\newcommand{\be}{\begin{equation}}
\newcommand{\bea}{\begin{eqnarray}}
\newcommand{\ee}{\end{equation}}
\newcommand{\eea}{\end{eqnarray}}

\newcommand{\qe}{\varepsilon}

\newcommand{\fr}[2]{{\textstyle \frac{#1}{#2}}}

\newcommand{\cX}{{\mathcal X}}

\documentclass{iet-ell}
\usepackage{array}

\usepackage{natbib}
\renewcommand\harvardurl[1]{\textbf{URL:} \url{#1}}
\usepackage{hyperref}

\setcopyright{open}
\ietVolume{00}
\ietYear{2021}
\ietDoi{ell.10001}

\received{DD MMMM YYYY}
\accepted{DD MMMM YYYY}

\begin{document}

\title{Implementation of Entropically Secure Encryption: Securing Personal Health Data}

\author[af1]{Mehmet H\"{u}seyin Temel}
\orcid{0000-0002-7386-6396}
\author[af2]{Boris \v{S}kori\'{c}}
\author[af1]{Idelfonso Tafur Monroy}

\affil[af1]{Electrical Engineering Department, Eindhoven University of Technology, Eindhoven, The Netherlands}
\affil[af2]{Mathematics and Computer Science Department, Eindhoven University of Technology, Eindhoven, The Netherlands}

\corresp{m.h.temel@tue.nl}

\begin{abstract}
Entropically Secure Encryption (ESE) offers 
unconditional security with shorter keys compared to the One-Time Pad.
In this paper, we present the first implementation of ESE for bulk encryption. 
The main computational bottleneck for bulk ESE is a multiplication in a very large finite field.
This involves multiplication of polynomials followed by modular reduction. 
We have implemented polynomial multiplication based on the {\tt gf2x} library,
with some modifications that avoid inputs of vastly different length, thus improving speed.
Additionally, we have implemented a recently proposed efficient reduction algorithm that works for any polynomial degree. We investigate two 
use cases: X-ray images of patients and human genome data. 
We conduct entropy estimation using compression methods whose results determine the key lengths required for ESE. 
We report running times for all steps of the encryption.
We discuss the potential of ESE to be used in conjunction with Quantum Key Distribution (QKD), in order to 
achieve full information-theoretic security of QKD-protected links for these use cases.
\end{abstract}

\maketitle

\section{Introduction}
Practical encryption schemes rely on computational hardness assumptions, 
making them potentially susceptible to advances in cryptanalysis. 
The One-Time Pad (OTP) encryption scheme \cite{Vernam1918} provides unconditional security, not depending on any assumptions;
however, the size of the key needs to equal at least the entropy of the message, making OTP-ing impractical. 
OTP encryption works by XOR-ing the key and the message.
OTP provides perfect security, which means that the ciphertext leaks no information about the message.

\subsection{Entropically Secure Encryption} 
If one does not aim for {\em perfect} security, it is possible to achieve unconditional security with shorter keys. 
The only additional condition is to guarantee a lower bound on the entropy of the message from the point of view of the attacker. 
This concept is called entropic security and the  encryption scheme is called entropically secure encryption (ESE) \cite{russell2002fool, dodis2005entropic} . 
If the plaintext has at least $t$ collision entropy, then it can be encrypted using a key with a length $\ell$ of 
\begin{equation}
    \ell = n-t+2\log\fr1\qe 
\label{keylength}
\end{equation}
where $n$ is the message length and $\qe$ is the security parameter \cite{dodis2005entropic, temel2024entropically}. 
This implies that a message whose collision entropy is close to its length (i.e. almost uniformly distributed)
requires only a short key for unconditional security.
The collision entropy of a random variable with probability mass function $(p_x)_{x\in\cX}$ is defined as 
$-\log_2 (\sum_{x\in\cX} p_x^2)$.

In ESE, a short key is expanded to a long pseudorandom string which is then used
as the key for OTP. The key expansion is accomplished through Galois Field (GF) multiplication of a public random string and the key. 
Although ESE has been well studied, there is no implementation for encrypting bulk data, possibly because of the following reasons. 
(1)~Entropy estimation of the plaintext. 
If the data is not generated according to a known probability distribution, it is impossible to compute the entropy precisely.
A way must be found to lower bound the entropy, which is not always straightforward.
(2)~Typical implementations of GF multiplication work with data sizes that are much smaller than in bulk ESE.

\section{Entropically Secure Encryption and QKD}
QKD has the unique property that it generates unconditionally secure keys.
However, the practical implementation of QKD often involves the subsequent use of {\em computationally} secure encryption methods such as AES to secure transmitted data \cite{alleaume2010quantum}. 
Herein lies an opportunity for further security improvement through the integration of ESE with QKD:
if the QKD system produces key bits at a rate that is compatible with the key consumption of the ESE
(which depends on the type of data and the transmission rate of the bulk channel) then 
full information-theoretic security of the communication channel can be achieved.
As the entropy condition for ESE is tricky, it is prudent to apply a layer of computationally secure
encryption as well.

\section{Key Expansion}
Key expansion consists of two steps: 
(1) binary polynomial multiplication of the public string X and the key $k$; followed by
(2) modular reduction of the result with an irreducible polynomial. 

\subsection{Binary Polynomial Multiplication}
Our goal is to achieve bulk encryption with relatively short keys. 
Consequently, the multiplication algorithm must be suitable for highly unbalanced multiplication scenarios, 
where a lengthy public string—possibly megabytes or gigabytes—is multiplied by a much shorter key. 

A number of software libraries is capable of handling operands larger than one~MB.
Chen et al. \cite{chen2017faster} presented the {\tt Add-FFT} algorithm, which 
takes two inputs of equal length up to $2^{36}$ bits.
At the time of this research, {\tt Add-FFT} is the fastest library for balanced (equal length)
multiplication beyond $2^{17}$ bits.
The {\tt gf2x} library \cite{brent2008faster}, one of the most common for GF operations, accepts arbitrary input lengths.
Other libraries exist for multiplication of binary polynomials, but they cannot handle the lengths with which we want to work.  

We benchmarked {\tt Add-FFT} and {\tt gf2x} for unbalanced input sizes. (An example is shown in Table~\ref{tab:gf-fft}).
For {\tt Add-FFT} we applied zero-padding of the short string in order to get equal lengths.
Our setup was a laptop with an Intel Core i7-9750H 2.60GHZ processor and 16GB RAM. 
For many different operand sizes we observe that 
{\tt gf2x} is much faster than {\tt Add-FFT} when the operands are unbalanced.
Hence we selected {\tt gf2x} for further experiments (see Table~\ref{tab:gf2x-k}).
We observe a sharp increase at key length $2^{10}$.
Due to this artefact
the running times are clearly too long for applications with a high data rate.
A new multiplication implementation becomes essential for achieving more efficient encryption.
\begin{table}[!h]
\centering
\caption{Execution times (in seconds) of the binary polynomial multiplication
for fixed key length $2^7$ bits and varying message length (in bits).}
\label{tab:gf-fft}
\begin{tabular}{rcccccc} 
\textbf{Length of X}& $2^{22}$ & $2^{24}$ & $2^{26}$ & $2^{28}$ & $2^{30}$ & $2^{32}$ \\
\hline
{\tt Add-FFT}  & 0.02& 0.08& 0.47& 2.60& 9.67& 40.77\\
{\tt gf2x}  & 0.0005 & 0.002 & 0.003 & 0.01& 0.05& 0.20\\
\end{tabular}
\end{table}

\begin{table}[!h]
\centering
\caption{Execution times of {\tt gf2x} binary polynomial multiplication at fixed message length $2^{30}$ bits and varying key length (in bits).  }
\label{tab:gf2x-k}
\begin{tabular}{rccccccc} 
\textbf{Length of k}& $2^{7}$& $2^{8}$& $2^{9}$& $2^{10}$& $2^{11}$& $2^{12}$& $2^{13}$\\ 
\hline
time (s) & 0.05& 0.10& 0.23& 5.31& 5.28& 5.27& 5.34\\ 
\end{tabular}
\end{table}

\subsubsection{Simplemult} 

In an attempt to remove the artefact,
we developed a method {\tt simplemult} that performs the multiplication piecewise using blocks that have the size of the key.
Let the key be represented as the polynomial $q(x)$.
Let the random seed be $p(x) = \sum_{i=0}^{N-1} p_i(x) x^{i(1+\deg q)}$ where $N=\frac{1+\deg p}{1+\deg q}$ and
the polynomials $p_i$ are of degree $\deg(q)$.
The multiplication is computed as
\begin{equation}
    p(x)\; q(x) = \sum_{i=0}^{N-1} p_i(x) q(x)\, x^{i(1+\deg q)}.
\end{equation}
We use {\tt gf2x} for the multiplication $p_i(x) q(x)$.  
We have benchmarked {\tt simplemult} against {\tt gf2x}, see Table \ref{tab:simplemult-X30}. 
We also implemented a multithread version {\tt t\textunderscore simplemult} and executed it with 12 threads.

\begin{table}[!h]
\centering
\caption{Execution times (in seconds) of the binary polynomial multiplication
for fixed message length $2^{30}$ bits and varying key length (in bits).}
\label{tab:simplemult-X30}
\begin{tabular}{r>{\centering\arraybackslash}p{0.04\linewidth}>{\centering\arraybackslash}p{0.04\linewidth}>{\centering\arraybackslash}p{0.04\linewidth}>{\centering\arraybackslash}p{0.04\linewidth}>{\centering\arraybackslash}p{0.04\linewidth}>{\centering\arraybackslash}p{0.04\linewidth}>{\centering\arraybackslash}p{0.04\linewidth}} 
\textbf{Length of k}& $2^{7}$& $2^{8}$& $2^{9}$& $2^{10}$& $2^{11}$& $2^{12}$& $2^{13}$\\ \hline
{\tt gf2x} & 0.05& 0.10& 0.23& 5.31& 5.28& 5.27& 5.34\\ 
{\tt simplemult}& 0.12& 0.08& 0.07& 0.14& 0.19& 0.28& 0.40\\ 
{\tt t\_simplemult }& 0.08& 0.07& 0.07& 0.07& 0.07& 0.07& 0.09\\ 
\end{tabular}
\end{table}
Table \ref{tab:simplemult-X30} demonstrates a significant superiority of our algorithm 
over {\tt gf2x} when the key length is greater than $2^{7}$. 
This improvement reaches more than 36-fold when the key length is $2^{10}$. 
When the key is smaller than $2^{8}$, our algorithm is slower because of overhead. 
Additionally, multithreading gives a further speedup.
This indicates that there is potential for further optimisation on different platforms and with different multithreading options.
We note that {\tt gf2x} itself cannot be sped up by multithreading since {\tt gf2x} is a sequential process.

We have done further benchmarks at fixed key length $2^{14}$ and varying message length (Table \ref{tab:simplemult-k14}).
Again, our method is significantly faster than {\tt gf2x}, and multithreading is improving performance.

\subsection{Reduction} 
The second step of the key expansion is modulo reduction of the multiplication result with an irreducible polynomial. 
The choice of this polynomial is specific to the message length.
Lists of irreducible polynomials are available up to degree  around $2^{14}$ \cite{seroussi1998table}.
Furthermore, some cryptographic primitives provide irreducible polynomials of very specific degrees \cite{rijmen2001advanced}.
However, we need a broad range of degrees.
We have implemented a class of irreducible polynomials and the reduction algorithm proposed by Banegas et al. \cite{banegas2019new}. 
The polynomials have flexible degree, and the reduction algorithm is efficient.
Table \ref{tab:simplemult-k14} lists running times of our implementation {\tt redBCP}.

\begin{table}[!h]
\centering
\caption{Execution times (in seconds) of the binary polynomial multiplication and reduction algorithms for fixed key length $2^{14}$ bits and varying message length (in bits).}
\label{tab:simplemult-k14}
\begin{tabular}{rccccc>{\centering\arraybackslash}p{0.05\linewidth}} 

\hline
\textbf{Length of X}& $2^{22}$ & $2^{24}$ & $2^{26}$ & $2^{28}$ & $2^{30}$ & $2^{32}$ \\
\hline
 & \multicolumn{6}{c}{\textbf{Multiplication Algorithms} } \\ \cline{2-7}
{\tt gf2x}& 0.01& 0.06& 0.29& 1.34& 5.17& 44.3\\

{\tt simplemult}& 0.002& 0.009& 0.04& 0.15& 0.58& 2.30\\

{\tt t\textunderscore simplemult}& 0.001& 0.004& 0.008& 0.03& 0.12& 0.47\\
\hline
 & \multicolumn{6}{c}{\textbf{Reduction Algorithms} } \\ \cline{2-7}
{\tt redBCP}& 0.03 & 0.12 & 0.48 & 1.94 & 8.32 & 34.87 \\

{\tt t\textunderscore redBCP}& 0.01 & 0.02 & 0.06 & 0.23 & 0.91 & 3.89 \\
\end{tabular}
\end{table}

Our reduction takes longer than the polynomial multiplication,
even though in theory it is an easier task.
The reason is that our implementation {\tt redBCP} is not in any way optimised.
We were aiming only for a functional implementation of \cite{banegas2019new}.
We expect that code optimisation will yield running times below {\tt simplemult}.
Note that 12-thread multithreading ({\tt t\textunderscore redBCP}) leads to a significant speedup,
up to a factor~9.

\section{Entropy Estimation} 
From (\ref{keylength}) it is clear that the key length depends on the characteristics of the plaintext,
in particular a lower bound $t$ on the collision entropy.
On the encryption side, such a lower bound has to be available. 
In practical scenarios the plaintext is not generated from a known fixed distribution,
and the entropy cannot be computed exactly.

It is well known that on average, a piece of data cannot be compressed below its Shannon entropy \cite{cover1999elements}. 
We take the approach of compressing data and then using the compressed size as an estimate
for the collision entropy.
It must be noted that collision entropy is {\em smaller} than Shannon entropy unless the distribution is entirely uniform; 
hence we obtain an overestimate even
if we get the Shannon entropy correct.
Nonetheless, due to lack of other methods we explore this approach.
We consider only lossless compression, since lossy compression destroys information, which would make our task more difficult.

We looked at two use cases and gathered real-world data for these cases. 
We benchmarked various compression methods suitable for the selected data and 
based our entropy estimates on the methods yielding the best compression. 
Then we used (\ref{keylength}) to estimate the required key length in a particular way:
(i) The value of $t$ follows from the above described approach of finding the best compression;
(ii) The value of $n$ is determined by the standard way in which the data is communicated 
in the use case. 
For instance, this can be a non-optimally compressed form dictated by a standard.
Then the key length is essentially the size of the badly compressed file minus the size of the optimally compressed file.

In the first scenario, we focused on chest X-ray images from the MIMIC-IV dataset \cite{johnson2020mimic, goldberger2000physiobank}, 
a widely used dataset. 
For the second use case, we used a human genome dataset of the study PRJEB36890 \cite{Embl-Ebi_2020} from the 1000 Genome Project \cite{10002015global}. 
The dataset comprises human DNA sequencing information, with an average size of 130 GB, 
aligning with our aim to do bulk encryption.
We selected these datasets because they contain a type of personal data that is highly sensitive. 

\subsubsection{X-Ray Image Use Case} 
The X-Ray image files are in DCM format, a standard for medical images that includes patient information \cite{DICOM}. 
DCM files follow the DICOM (Digital Imaging and Communications in Medicine) format.
For our analysis, we utilized a sample of 881 files from the dataset, 10.7 GB, with an average file size of approximately 12.5 MB. 
We benchmarked various compression methods. 
We present the compression ratios of the top~6 (Table~\ref{tab:Xray-comp}). 
BZip2 and LZMA2 are universal methods with generic applications, while the other four methods are versions of general methods implemented in the GDCM \cite{GDCM_2019} application following the DICOM format.
\begin{table}[!h]
\caption{Compression ratios for X-ray images. Ratio: compressed size/uncompressed size.}
\label{tab:Xray-comp}
\begin{tabular}{rcccccc} 
\hline
\textbf{Method} & BZIP2 & LZMA2 & RLE& JPEG& J2K& JPEG-LS\\ \hline
Ratio & 0.37 & 0.43 & 0.56 & 0.44 & 0.40 & 0.40 \\ 
\end{tabular}
\end{table}

Based on the findings from Table~\ref{tab:Xray-comp}, we choose BZIP2's compression results for our entropy estimation. 
On the other hand, DICOM mandates one of the standard methods; since JPEG-LS is the best-performing standard method we use JPEG-LS
compressed files as the data that needs to be encrypted.
The average JPEG-LS-compressed file size is 5 MB, while the average estimated Shannon entropy for a file is approximately 4.62~MB. 
The resulting key length~(\ref{keylength}) for ESE-encrypting a JPEG-LS-compressed
X-ray file is around 0.38 MB.
Compared to OTP encryption, which needs a 5 MB key, that is a reduction of 93\%.

\subsubsection{Human DNA Use Case}  
We used a sample of approximately 8 TB from the 1000 Genome Project dataset, comprising 61 files with an average size of 132.12 GB.  
In Table \ref{tab:Dna-comp} we present the four best methods. 
The first three are well known general methods. 
Spring \cite{chandak2019spring} is a tool specifically developed for processing and compressing genome files. 
With Spring, the average compressed file size is around 6.49 GB.  
It is worth noting that some DNA compression methods claim to provide better ratios, but since they are lossy \cite{hernaez2019genomic},
we do not consider them our analysis.
We select the Spring compression results as our entropy estimation.
\begin{table}[!h]
\caption{Compression ratios for genome files. Ratio: compressed size/uncompressed size.}
\label{tab:Dna-comp}
\begin{tabular}{rcccc}
\hline
\textbf{Method} & BZIP2 & LZMA2 & PPMD   & SPRING\\ \hline
Ratio& 0.083 & 0.075 & 0.063 & 0.049 
\end{tabular}
\end{table}

If Spring-compression were ideal, yielding a completely uniform output, then 
the ESE key length would be $2\log\frac1\qe$, which is tiny compared to the file size.
However, it is too optimistic to assume that Spring produces ideal compression.
We want to quantify the gap between the compressed size and the entropy.
Unfortunately there is no obvious way to achieve this.
We present a heuristic method to estimate the gap.
We notice that not all files are compressed at the same ratio;
we observe a standard deviation of around 0.006 in the compression ratio.
Handwavingly we postulate that optimal compression occurs at the average ratio minus one standard deviation,
i.e. $0.049 - 0.006 =0.043$.
For an average file size of 132.12 GB this yields an entropy of 5.68 GB
and a compressed size of 6.47 GB.
The resulting gap, and hence key length, is around 800~MB.
That is not a large relative reduction compared to the compressed file size of 6.47 GB.
Note however that our heuristic argument could be too pessimistic;
it assumes that all genome files have the same information density, which may be false.

\section{Complete Implementation} 
We have implemented the entire ESE scheme by integrating our multiplication algorithm {\tt simplemult}, 
the {\tt t\textunderscore redBCP} reduction, and the final XOR operation of the message and the expanded key. 
We ran experiments
on two setups: 
(i) a personal laptop with Intel Core i7-9750H 2.60GHz(6 cores) with 16GB RAM and 
(ii) a server with Intel Xeon Silver 4214R 2.40GHz (48 cores) with 376 GB RAM. 
We set $\qe=2^{-128}$, which gives $2\log\frac1\qe=256$.
(Given the overall key sizes, the choice of $\qe$ has a negligible effect.)
We present execution times in Tables \ref{tab:Xray-enc}, \ref{tab:exp-49} and \ref{tab:exp-43}.

In our experiment with X-ray images, files are encrypted as a whole because 
this is feasible for the average compressed file size of 5MB.

In the genome use case, the average compressed file size is around 6.49 GB which makes it difficult to encrypt 
an entire file at one time. 
Therefore, we divided the files into smaller chunks and encrypted them separately. 
This increases the key length by 256 bits for every additional chunk due to the 
$2\log\frac1\qe$ term in (\ref{keylength}), which applies to each individual encryption .

\begin{table}[!h]
\centering
\caption{Average execution time (in seconds) of the sub-processes in the encryption of one X-ray image file. 
The encryption rate and key consumption rates are in MB/s.}
\label{tab:Xray-enc}
\begin{tabular}{rccccc} 
\hline
 & \textbf{Mult}.& \textbf{Red}.& \textbf{XOR} & \textbf{Enc. Rate}& \textbf{Key Cons.}\\ \hline
Laptop & 0.22 & 0.05 & 0.0010 & 18.79 & 1.42 \\ 
Server    & 0.22 & 0.03 & 0.0005 & 19.83 & 1.50 \\ 
\end{tabular}
\end{table}

In our first experiment with genome files (Table~\ref{tab:exp-49}), we assume that compression is optimal, yielding a key length 256 bits per chunk. 
This leads to very small key consumption compared to the file size.
\begin{table}[!h]
\centering
\caption{Average execution times (in seconds) of the sub-processes of the ESE scheme for one genome file with various chunk sizes (MB) 
when the key length is 256 bits per chunk. 
The encryption rate is in MB/s and key consumption is in bits/s.}
\label{tab:exp-49}
\begin{tabular}{cccccc} 
\hline
\textbf{Chunk Size}& \textbf{Mult.} & \textbf{Red.} & \textbf{XOR} & \textbf{Enc. Rate}& \textbf{Key Cons.}\\
\hline
 &\multicolumn{5}{c}{\textbf{Laptop}}\\ \cline{2-6}
256 & 16.99 & 68.52 & 1.50 & 76.37 & 79.44 \\
512 & 17.20 & 71.79 & 2.07 & 72.97 & 39.36 \\
1024 & 21.84 & 74.93 & 2.53 & 66.92 & 18.05 \\
 &\multicolumn{5}{c}{\textbf{Server}}\\ \cline{2-6}
256 & 15.67 & 48.43& 1.39& 101.47& 106\\
512 & 15.76& 48.46& 1.50& 101.12& 55\\
1024 & 15.85& 49.47& 1.67& 99.20& 27\\
2048 & 16.13& 50.45& 1.89& 97.06& 15\\
4096 & 17.16& 52.48& 2.46& 92.16& 7\\
\end{tabular}
\end{table}

In the second experiment (Table \ref{tab:exp-43}), we assume that the entropy ratio in a genome file is 0.043, 
which is lower than the compression ratio~0.049. 
This yields a key length (per chunk) given by
$\approx \frac{\rm chunk size}{0.049}(0.049 - 0.043)$.

\begin{table}[!h]
\centering
\caption{Average execution times (in seconds) of sub-processes of the ESE scheme on the server for one genome file with various chunk sizes (MB) assuming entropy ratio 0.043. Key length (MB) is per chunk. Encryption and key consumption rates are in MB/s.}
\label{tab:exp-43}
\begin{tabular}{>{\centering\arraybackslash}p{0.1\linewidth}>{\centering\arraybackslash}p{0.1\linewidth}ccc>{\centering\arraybackslash}p{0.1\linewidth}>{\centering\arraybackslash}p{0.1\linewidth}} 
\hline
\textbf{Chunk Size}&  \textbf{Key Length}&\textbf{Mult.} & \textbf{Red.} & \textbf{XOR} & \textbf{Enc. Rate}& \textbf{Key Cons.}\\ \hline
256 &  31.34&798& 50.33& 1.39& 7.82& 0.94\\
512 &  62.70&627& 51.74& 1.50& 9.77& 1.18\\
1024 &  125.39&792& 55.12& 1.67& 7.82& 0.94\\
2048 &  250.78&1322& 53.92& 1.89& 4.82& 0.58\\
\end{tabular}
\end{table}

While in Table \ref{tab:exp-49} the encryption rate decreases with increasing chunk size, in Table \ref{tab:exp-43}, the encryption rate reaches its peak with a chunk size of 512 MB. 
This is due to the trade-off between the lengths of the operands and the number of chunks in {\tt simplemult}.  
Thus for the genome use case, the optimal chunk size to be encrypted is around 512 MB.

\section{Conclusion}
We have implemented all stages of Entropically Secure Encryption and studied 
execution times for two use cases.
The computational bottleneck is the polynomial multiplication.
We see room for improvement by better use of parallelization.
Furthermore, the implementation of the reduction can be significantly improved.

In the X-ray images use case, we get file encryption in around 0.3 seconds, with a key consumption rate of
1.5 MB/s.
In the DNA use case with a heuristic entropy estimate, which may be highly pessimistic, 
we get optimal performance when we independently encrypt chunks of 512~MB,
with a key consumption rate around 1 MB/s.
With an optimistic estimate of the compression efficiency we get much smaller keys,
leading to increased encryption speed and key consumption of less than 100~bits/s.

For the use of ESE in conjunction with QKD, the key generation rate of QKD must exceed the 
speed at which key material is used up, which is dictated by the data rate of the communication channel and the ESE key size
associated with the data type.
For simplicity, we consider the case that compressed files need to be sent at 10 MB/s which is a reasonable rate for a hospital \cite{rateguideFCC} .
In the X-ray use case, this implies that there is 0.5s available to encrypt a compressed file,
which is feasible on a laptop.
Furthermore, it leads to a key consumption of 0.6 MB/s.
In the genome use case with pessimistic assumptions (Table~\ref{tab:exp-43})
and 512MB chunk size, the encryption rate struggles to match the data rate,
and the key consumption is 1.2 MB/s.

Whether the key consumption matches the QKD key generation rate
depends on the use case, the deployed QKD technology and the QKD distance. 
Current QKD implementations achieve key rates up to approximately 14~MB/s over distances of about 10 km \cite{li2023high, bacco2019boosting}
and 0.4~MB/s at around 100 km \cite{grunenfelder2023fast}.  
At distances up to roughly 50 km this is certainly compatible with the key consumption in both our use cases,
even under pessimistic assumptions.
Beyond that range, it is certainly compatible with the genome use case under mildly optimistic assumptions on
the quality of the compression.

In conclusion, even an un-optimised implementation of Entropically Secure Encryption 
shows that ESE should be practical in the two studied use cases, and that 
integration with QKD is feasible given the matching key rates.
As topics for future work we highlight
further optimization of the finite-field multiplication,
studying more use cases,
and improved entropy estimates.

\begin{acks}
Part of this work was supported by the Dutch Startimpuls NAQT CAT-2 and NGF Quantum Delta NL CAT-2
\end{acks}

\bibliography{iet-ell}
\bibliographystyle{iet}

\end{document}